\documentclass[11pt]{article}
\usepackage{amsfonts}
\usepackage{amsmath,amssymb}

\newcommand{\om}{\Omega}

\newcommand{\bec}{\begin{center}}
\newcommand{\eec}{\end{center}}

\newcommand{\bea}{\begin{array}}
\newcommand{\ear}{\end{array}}

\newcommand{\bfr}{\begin{flushright}}

\newcommand{\efr}{\end{flushright}}
\newcommand{\noi}{\noindent}
\newcommand{\me}{\frac{1}{2}}

\newcommand{\cl}{{\mt{C}}\ell}

\newcommand{\RR}{\mathbb{R}}\newcommand{\op}{\oplus}
\newcommand{\HH}{\mathbb{H}}\newcommand{\PP}{\mathbb{P}}
\newcommand{\ap}{\alpha}

\newcommand{\ot}{\otimes}
\newcommand{\la}{\Lambda}
\newcommand{\bege}{\begin{equation}}
\newcommand{\enge}{\end{equation}}

\newcommand{\g}{\gamma}
\newcommand{\ri}{\rightarrow}

\newcommand{\ty}{\RR\oplus\RR^3}
\newcommand{\si}{\sigma}

\newcommand{\beq}{\begin{eqnarray}}\newcommand{\benu}{\begin{enumerate}}\newcommand{\enu}{\end{enumerate}}
\newcommand{\eeq}{\end{eqnarray}}
\newcommand{\mt}{\mathcal}

\newcommand{\ee}{{\bf e}}

\newcommand{\mr}{\mathring}

\newcommand{\CC}{\mathbb{C}}

\newcommand{\ZZ}{\mathbb{Z}}

\newcommand{\ol}{\overline}

\newcommand{\clt}{{\mt{C}}\ell_{3,0}}
\newcommand{\cle}{{\mt{C}}\ell_{1,3}}

\newcommand{\xx}{{\bf x}}

\newcommand{\BA}{\breve{A}}

\newcommand{\mmu}{{\mathfrak{u}}}

\newcommand{\bx}{\begin{pmatrix}}
\newcommand{\ex}{\end{pmatrix}}
\newcommand{\vcx}{\varepsilon}

\newcommand{\mmb}{{\mathfrak{b}}}
\newcommand{\mmg}{{\mathfrak{g}}}

\usepackage{amsfonts}
\usepackage{amsmath,amssymb}

\usepackage{indentfirst}  
\begin{document}
\title{Twistors, Generalizations and Exceptional Structures}
\author{{\bf Rold\~ao da Rocha}\thanks{Instituto de F\'{\i}sica Gleb Wataghin (IFGW), Unicamp, CP 6165, 13083-970, Campinas (SP), Brazil. 
E-mail: roldao@ifi.unicamp.br. Supported by CAPES.}
\and{\bf Jayme Vaz, Jr.}\thanks{Departamento de Matem\'atica Aplicada, IMECC, Unicamp, CP 6065, 13083-859, Campinas (SP), Brazil. E-mail: 
vaz@ime.unicamp.br}}

\date{}\maketitle

\abstract${}^{}$\begin{center}
\begin{minipage}{12cm}$ \;\;\;\;\;$
{This paper is intended to describe twistors via the paravector model of Clifford algebras
and to relate such description to conformal maps in the Clifford algebra over $\RR^{4,1}$, besides pointing out some applications
of the pure spinor formalism. We construct
 twistors in Minkowski spacetime as algebraic spinors associated with the Dirac-Clifford algebra
 $\mathbb{C}\otimes C\ell_{1,3}$ 
 using one lower spacetime dimension than standard Clifford algebra formulations, since for this purpose 
 the Clifford algebra over $\RR^{4,1}$ is also used to describe conformal maps, instead of $\RR^{2,4}$.
It is possible to identify, via the pure spinor formalism,
 the twistor fiber in four, six and eight 
dimensions, respectively, with the coset spaces ${{\rm SO(4)}}/({\rm SU(2)\times U(1)/\ZZ_2)}\simeq\CC\PP^1$, ${\rm SO(6)}/({\rm SU(3)\times
 U(1)/\ZZ_2)}\simeq \CC\PP^3$ and ${\rm SO(8)}/({\rm Spin(6)\times Spin(2)/\ZZ_2)}$.
The last homogeneous space is closely related to the SO(8) spinor decomposition preserving SO(8) symmetry in type IIB superstring theory.
Indeed, aside the IIB superstring theory, 
there is no SO(8) spinor decomposition preserving SO(8) symmetry and, in this 
case,
 one can introduce distinct coordinates and conjugate momenta only if the Spin(8) symmetry is broken by a Spin(6)$\times$ Spin(2)
 subgroup of Spin(8).  Also,
it is reviewed how to generalize the Penrose flagpole, illustrating
 the use of the pure spinor formalism to construct a flagpole that is more general than 
the Penrose one, which arises when a defined parameter goes to zero.
We investigate the well-known relation between this flagpole and the  SO($2n$)/U($n$) twistorial structure,
which emerges when one considers the action of a suitable classical group on the set $\Xi$ of all totally
isotropic subspaces of $\CC^{2n}$, and an isomorphism from the set of pure spinors to $\Xi$.
We follow the F. Reese Harvey's book approach.
Finally we point out some relations between twistors fibrations and the classification
of compact homogeneous
quaternionic-K\"ahler manifolds (the so-called Wolf spaces), and exceptional Lie structures, whose grading 
is also briefly commented.}
\end{minipage}\end{center}
\medbreak
\medbreak\noi
%\FullConference{Fourth International Winter Conference on Mathematical Methods
%in Physics\\
%		 09 - 13 August 2004\\
	%	 Centro Brasileiro de Pesquisas Fisicas (CBPF/MCT), Rio de Janeiro, Brazil}

\section{Introduction}

Nowadays the search for any  unified theory that describes the four fundamental interactions demands a deep mathematical background
and an interface between physics and mathematics. 
The relation between superstring theory in twistor spaces \cite{mot,b1}  and the pure spinor formalism \cite{cartan,chev}
has been increasingly and widely investigated \cite{b2,b4}. 
 With the motivation concerning the SO(8) spinor decomposition that preserves SO(8) symmetry in type IIB superstring theory \cite{grscwi}, 
among others,
it can be shown via the pure spinor formalism the well-known result asserting that a twistor in eight dimensions is an element of
 the homogeneous space  SO(8)/(Spin(6)$\times$ Spin(2)/$\ZZ_2) \simeq$ SO(8)/U(4), and, in $n$ dimensions, an element of SO($2n$)/U($n$).

The main aim of this paper, besides  pointing out some relation between twistors and pure spinors, is to 
describe conformal maps in Minkowski spacetime as the twisted adjoint representation of  \$pin$_+$(2,4) 
(to be precisely defined in Sec. 2)
 on paravectors\footnote{A paravector of the Clifford algebra $\cl_{p,q}$ is an element of $\RR\op\RR^{p,q}$.} \cite{baylis,port}
 of $\cl_{4,1}$, and to characterize twistors as algebraic spinors\footnote{Algebraic spinors are 
 elements of a minimal lateral ideal 
of a Clifford algebra.} \cite{chev} in $\RR^{4,1}$. 
Although some papers have already described twistors using the algebra $\CC\otimes\cl_{1,3} \simeq \cl_{4,1}$ \cite{crau,abla,ke97}, the present formulation
sheds some new light on the use of the paravector model.
% In this way, it is possible   
%to generalize twistors to any even dimension.
 This paper is presented as follows: in Sec. 2 we describe conformal transformations using 
the twisted adjoint representation of the group SU(2,2) $\simeq\$$pin$_+$(2,4) on paravectors of $\cl_{4,1}$.
In Sec. 3 twistors, the incidence relation between twistors and the Robinson congruence,
 via multivectors and the paravector model of $\CC\ot\cle\simeq\cl_{4,1}$, are introduced. We show explicitly how our
 results can be led to the well-established ones of Keller \cite{ke97},
and consequently to the classical formulation introduced by Penrose \cite{pe1,pe2}. 
It is also described how one can obtain twistors as elements of SO($2n$)/U($n$) via pure spinors.
Finally in Sec. 4 we link twistor theory to Lie exceptional structures. 
\section{Conformal compactification and the paravector model}
Given a vector space, endowed with a metric $g$  of signature $p-q$, and denoted by
 $\RR^{p,q}$, consider the injective map  \cite{port}
                $\RR^{p,q}\ni   x\mapsto  (x, g(x,x), 1) = (x, \lambda, \mu)\in \RR^{p+1,q+1}$.            
The image of $\RR^{p,q}$ under this map is a subset of the Klein absolute 
$x\cdot x - \lambda\mu = 0$. 
This map induces an injective map from the conformal compactification $(S^{p}\times S^q)/\ZZ_2$ of $\RR^{p,q}$ to the projective space
$\RR\PP^{p+1,q+1}$.

The conformal group ${\rm Conf}(p,q)$ 
is isomorphic to the quotient group ${\rm O}(p+1,q+1)/\ZZ_2$ \cite{port}, and since the group ${\rm O}(p+1,q+1)$ has four components,
 then Conf$(p,q)$ has two (if $p$ or $q$ are even) or four components (otherwise) \cite{port,cru}.
Taking the case when $p=1$ and $q=3$, the group Conf(1,3) has four components, and the component  Conf$_+(1,3)$ connected to the identity
is the M\"obius group\footnote{All M\"obius maps are composition of rotations, translations, dilations and inversions
 \cite{maks}.} of $\RR^{1,3}$. Besides, the orthochronous connected component is denoted by SConf$_+$(1,3). 
Consider a basis
%\footnote{The elements of this basis satisfy the relations $\vcx_0^2 = \vcx_5^2 = 1,\; \vcx_1^2 =
% \vcx_2^2 = \vcx_3^2 = \vcx_4^2 = -1$ and $\vcx_{\BA} \cdot \vcx_{\BB} = 
%0\quad(\BA\neq\BB).$}
  $\{\varepsilon_{\BA}\}_{\BA = 0}^5$ of $\RR^{2,4}$
and  a basis
%\footnote{The elements of this basis satisfy the relations $E_0^2 = -1,\; E_1^2 = E_2^2 = E_3^2 = E_4^2 = 1$
%and  $E_A\cdot E_B = 0 \quad (A\neq B).$}
 $\{E_A\}_{A = 0}^4$ of $\RR^{4,1}$. This last basis can be obtained
 from $\{\vcx_{\BA}\}$ if the isomorphism $E_A \mapsto \vcx_{A}\vcx_5$ 
is defined. 

Given $\phi$ an element of the Clifford algebra $\cl_{p,q}$ over  $\RR^{p,q}$,  the reversion 
of $\phi$ is defined and denoted by $\tilde{\phi} = (-1)^{[k/2]}\phi$ ([$k$] expresses the integer part of $k$),
  while the graded involution acting on $\phi$  is defined by 
$\hat{\phi} = (-1)^k \phi$. The Clifford conjugation $\bar\phi$ of $\phi$ 
is given by the reversion composed with the main automorphism.

If we take a vector $\ap = \ap^{\BA}\vcx_{\BA} \in \RR^{2,4}$, a paravector  $\mmb\in\RR\oplus\RR^{4,1}\hookrightarrow\cl_{4,1}$
can be obtained as $\mmb = \ap\vcx_5 = \ap^AE_A + \ap^5.$ 
From the periodicity theorem\footnote{The periodicity theorem of Clifford algebras asserts that $\cl_{p+1,q+1}\simeq\cl_{1,1}\ot\cl_{p,q}$. }
 \cite{atia} we have the isomorphism $\cl_{4,1}\simeq\cl_{1,1}\otimes\clt \simeq M(2,\CC)\ot\clt$, where $M(2,\CC)$ denotes the group of $2\times 2$ matrices with
complex entries. For i = 1,2,3 the isomorphism 
from $\cl_{4,1}$ to $\cl_{3,0}$ is given explicitly by $E_i \mapsto E_iE_0E_4 := \ee_i$, where $\{\ee_i\}$ denotes a basis
of $\RR^3$. Defining 
$E_\pm := \me(E_4 \pm E_0)$, we can write $\mmb = \ap^5 + (\ap^0 + \ap^4)E_+ + (\ap^4 - \ap^0) E_- + \ap^i\ee_iE_4E_0$,
and then it is possible, if we represent $E_+ =  {{\left(\bea{cc}
              0&0\\1&0\ear\right)}}$ and $ E_- = {{ \left(\bea{cc}
              0&1\\0&0\ear\right)}},$ to write $\mmb =  {\footnotesize{\left(\bea{cc}
              \ap^5 + \ap^i\ee_i&\ap^4 - \ap^0\\\ap^0 + \ap^4&\ap^5 - \ap^i\ee_i\ear\right)}}$.                              
            The vector $\ap\in\RR^{2,4}$ is in the Klein absolute, and so  $\ap^2 = 0$.
 Besides, we assert that $\mmb$ is in the Klein absolute
if and only if $\ap$ is. Indeed, denoting  $ \lambda = \ap^4 - \ap^0$ and $\mu = \ap^4 + \ap^0$, 
if $\mmb{\bar{\mmb}} = 0$,  
the matrix element $(\mmb{\bar{\mmb}})_{11}$ is given by 
 \bege \label{27}             
 (\mmb{\bar{\mmb}})_{11} = x{\ol{x}} -\lambda\mu = 0, 
 \enge
 \noi where $x:= (\ap^5 + \ap^i\ee_i)\in \RR\oplus\RR^3\hookrightarrow \cl_{3,0}$. 
Choosing $\mu = 1$ then $\lambda = x{\ol{x}}$, and
this choice is responsible for a projective description. Also, 
the paravector $\mmb\in\RR\oplus\RR^{4,1}$ can be rewritten as
 $\mmb =  {\footnotesize{\left(\bea{cc}x &x{\ol x}\\
 1&{\ol{x}}
 \ear\right)}}.$
 From eq.(\ref{27}) we obtain $
  (\ap^5 + \ap^i\ee_i)(\ap^5 - \ap^i\ee_i) = (\ap^4 - \ap^0)(\ap^4 + \ap^0)$
 from where $(\ap^5)^2 + (\ap^0)^2 - (\ap^1)^2 - (\ap^2)^2 - (\ap^3)^2 - (\ap^4)^2 = 0,$ showing that $\ap$ is
indeed in the Klein absolute. 
 
 Now consider an element  $ g\in$ SU(2,2) $\simeq\${\rm pin}_+(2,4) := \{g \in \cl_{4,1} \;|\;g\ol{g} = 1\}$. 
From the periodicity theorem, it can be represented as 
$  g = {\footnotesize{ \left(\bea{cc} a&c\\
 b&d
 \ear\right)}}$, where $a,b,c,d\in \clt$. 
 
In order to perform a rotation of the paravector $\mmb$, 
we can use the twisted adjoint representation ${\hat{\si}}:\${\rm pin}_+(2,4) \ri {\rm SO}_+(2,4)$, defined by its 
action on paravectors by  
 ${\hat{\si}}(g)(\mmb) = g\mmb{\hat{g}}^{-1} = g\mmb{\tilde{g}}$. In terms of matrix representations (with entries in $\clt$), 
the group \$pin$_+$(2,4) acts on paravectors $\mmb$ 
as  $ g\mmb\tilde{g} = {\footnotesize{\left(\bea{cc} a&c\\
 b&d
 \ear\right) \left(\bea{cc} x&\lambda\\
 \mu&{\ol{x}}
 \ear\right)\left(\bea{cc} {\ol d}&{\ol c}\\
 {\ol b}&{\ol a}
 \ear\right)}}$.  Fixing $\mu = 1$, $\mmb$ is mapped on 
${\footnotesize{
\left(\bea{cc} a&c\\
 b&d
 \ear\right) \left(\bea{cc} x&x{\ol{x}}\\
 1&{\ol{x}}
 \ear\right)\left(\bea{cc} {\ol d}&{\ol c}\\
 {\ol b}&{\ol a}
 \ear\right) = \Delta\left(\bea{cc} x^\prime& x'{\ol{x}'}\\
 1&{\ol{x}'}
 \ear\right)}},$ where $x' = (ax + c)(bx + d)^{-1}\in\RR\oplus\RR^3$ and $
 \Delta = (bx + d)({\ol{bx + d}})\in\RR.$
In this sense the spacetime conformal maps are rotations in $\RR\op\RR^{4,1}$, 
performed by the twisted adjoint representation, just given above.   
 All the spacetime conformal maps are expressed  respectively by the following matrices \cite{port,maks,hes}: 
\begin{center}
\begin{tabular}{||r||r||r||}\hline\hline
Conformal Map&Explicit Map&Matrix of $\$$pin$_+(2,4)$\\\hline\hline
Translation&$x\mapsto x + h,\;\; h\in \RR\op\RR^3$ &{\footnotesize{$\left(\bea{cc} 1&h\\
 0&1
 \ear\right)$}}\\\hline
Dilation&$x\mapsto \rho x, \;\rho\in\RR $& {\footnotesize{$
\left(\bea{cc} \sqrt{\rho}&0\\
 0&1/\sqrt{\rho}
 \ear\right)$}}\\\hline
Rotation&$x\mapsto \mmg x {\hat{\mmg}}^{-1},\; \mmg\in \${\rm pin}_+(1,3)$ &{\footnotesize{
$\left(\bea{cc} \mmg&0\\
0&{\hat{\mmg}}
 \ear\right)$}}\\\hline
Inversion&$x\mapsto -{\ol{x}}$& 
{\footnotesize{
$\left(\bea{cc} 0&-1\\
1&0
 \ear\right)$}}\\\hline
Transvection& $x\mapsto x + x(hx + 1)^{-1},\;\;h\in\ty$&  {\footnotesize{
$\left(\bea{cc} 1&0\\
h&1
 \ear\right) $}} \\\hline\hline
\end{tabular}                   
\end{center}
\noi This index-free algebraic formulation allows to trivially generalize the conformal maps of $\RR^{1,3}$ to the ones of $\RR^{p,q}$,
 if the periodicity theorem of Clifford algebras is used. 
The homomorphisms $\${\rm pin}_+(2,4)\simeq$ SU(2,2) $\stackrel{2-1}{\longrightarrow}{\rm SO}_+(2,4) \stackrel{2-1}{\longrightarrow}
{\rm SConf}_+(1,3)$ are explicitly constructed in \cite{lau}.

The generators of  Conf(1,3)  are expressed, using a basis $\{\g_\mu\}\in\cle$ and denoting
 the volume element of $\RR^{1,3}$ by $\g_5 =\g_0\g_1\g_2\g_3$, as $P_\mu = \me(\g_{\mu} + i\g_\mu\g_5),\;$  
$K_\mu = -\me (\g_{\mu} - i\g_\mu\g_5),\; D = \me i \g_5,$ and
$M_{\mu\nu} = \me(\g_\nu\wedge\g_\mu).$ They satisfy the commuting relations
\beq
[P_\mu, P_\nu] &=& 0,\quad\quad [K_\mu, K_\nu] = 0,\quad\quad [M_{\mu\nu}, D] = 0,\nonumber\\  
 \left[M_{\mu\nu}, P_{\lambda}\right] &=& -(g_{\mu\lambda}P_\nu - g_{\nu\lambda}P_\mu),\quad 
\left[M_{\mu\nu}, K_{\lambda}\right] = -(g_{\mu\lambda}K_\nu - g_{\nu\lambda}K_\mu),\nonumber\\
 \left[M_{\mu\nu}, M_{\sigma\rho}\right] &=& g_{\mu\rho}M_{\nu\sigma} + g_{\nu\si}M_{\mu\rho} - g_{\mu\si}M_{\nu\rho} - g_{\nu\rho}M_{\mu\si},\nonumber\\ 
 \left[P_\mu, K_\nu\right] &=& 2(g_{\mu\nu} D - M_{\mu\nu}),\quad 
 \left[P_\mu, D\right] = P_\mu,\qquad
 \left[K_\mu, D\right] = -K_\mu,
\eeq \noi which are invariant under $P_\mu \mapsto - K_\mu$, $K_\mu \mapsto - P_\mu$ and $D\mapsto -D$.
\section{Twistors as geometric multivectorial elements}
In this section we present and discuss the construction of twistors as algebraic spinors of $\cl_{4,1}$, using the paravector model,
 and 
as elements of SO($2n$)/U($n$), via the pure spinor formalism. We first present the Keller's approach of twistor theory from Clifford algebras,
and then we present and discuss our equivalent construction, which is led to the Keller one, and finally the way how these two (Clifford algebra, index-free)
approaches can be led to the Penrose twistor definition.
\subsection{Twistors as algebraic spinors using the paravector model}
\label{Keller}
The {\it reference twistor} $\eta_\xx$ is defined \cite{ke97}, given $\xx\in\RR^{1,3}$ and 
a dotted covariant Weyl spinor\footnote{A Weyl spinor can always be written 
as $\frac{1}{2}(1 \pm i\g_5)\psi$, where $\psi$ is a Dirac spinor.} (DCWS) 
 $\Pi = \frac{1}{2}(1 - i\g_5)\psi = ({0, \xi})^t$, as the multivector
\bege
\eta_\xx = (1 + \g_5\xx)\Pi.
\enge 
 The above expression is an index-free geometric algebra version of Penrose twistor in $\RR^{1,3}$,
 since if a suitable representation\footnote{As Keller \cite{ke97}, we choose to use a representation that  differs from the Weyl representation
by a sign on the matrices representing $\g_1, \g_2$ and $\g_3$.}
of $\CC\ot\cl_{1,3}$ is used, we have
 \bege\label{twistor} \eta_\xx = (1 + \g_5\xx)\Pi = {{\left[\left(\bea{cc}
           i_2&0\\
           0&i_2\ear\right) + \left(\bea{cc}
           -i_2&0\\
           0&i_2\ear\right)\left(\bea{cc}
           0&{\vec{x}}\\
           {\vec{x}}^c&0\ear\right)\right]{0\choose\xi} = {i {\vec{x}}\xi\choose\xi}}},
\enge\noi where  
           $
           {\vec{x}} = {\footnotesize{\left(\bea{cc}
           x^0 + x^3 & x^1 + ix^2\\
           x^1 - ix^2 & x^0 - x^3\ear\right)}}$. The symbol $\vec{x}^c$ denotes the  $\HH$-conjugation of $x$
 and $i_2 := i\,\mathbf{1}_{2\times 2}$.  
 
The {adjoint Dirac spinor} is defined as ${\mr\psi}= \psi^\dagger\g_0 = ({\bar{\psi}}_1, {\bar{\psi}}_2, {\bar{\psi}}_3, {\bar{\psi}}_4)$
and the {transposed twistor} as
${ {\mr\eta}}_{ \xx} = {\mr\psi} \frac{1}{2}(1 + i\g_5)(1 + \g_5{\bar \xx}) = {\mr\Pi}( 1 + \g_5{\bar \xx}).
$
The scalar product ${ {\mr{\eta}}}_\xx\eta_\xx$ represents the expected value of $\g_5\xx$ 
with respect to the spinor $\Pi$, 
since ${ {\mr{\eta}}}_\xx\eta_\xx = {\mr\Pi}\Pi + 2{\mr\Pi}\g_5\xx\Pi + \xx^2{\mr\Pi}\Pi =  2{\mr\Pi} \g_5\xx\Pi.$
The tensor product $\eta_\xx {\mr\Pi} = (1 + \g_5 \xx) \Pi{\mr\Pi} = (1 + \g_5 \xx)q$, where
 $q = \Pi{\mr\Pi}$ is the chiral positive projection of the timelike vector $Q = \psi{\mr\psi}$,
is also presented \cite{ke97}.
It allows to interpret the relation between a twistor, a timelike vector $q$ and the flagpole $\g_5\xx q$,
given by the following multivector:  
\bege
\zeta_\xx := \eta_\xx {\mr\Pi} = (1 + \g_5 \xx)q = q + \g_5 \xx q = (1 - i \xx)q \in  \cl_{4,1}.
\enge
The incidence relation,
 that determines a point in spacetime from the intersection of two twistors is defined, leading to the Penrose description \cite{pe1,pe2},
 as
\bege\label{jj}
J_{\xx\xx} := {{\ol{\eta}}_\xx}\eta_\xx = {\mr{\Pi}}\g_5(\xx - \xx)\Pi = 0.
\enge
\noi The product $J_{\xx\xx}$ is invariant if $\eta_\xx$ is multiplied by a complex number. Then
 eight dimensions are reduced to six, 
which leads to the classical interpretation of a twistor related to the space $\CC\PP^3 \simeq$ SO(6)/(SU(3)$\times$ U(1)/$\ZZ_2$) \cite{b4,pe1,pe2,harv}. 

Keller  presents another inner product \cite{ke97}, corresponding to the same twistor, but relating distinct points in spacetime, as
$J_{\xx\xx'} = {{\ol{\eta}_\xx}}\eta_{\xx'} = {\mr{\Pi}}\g_5(\xx - \xx')\Pi.$
 This product is null if and only if $\xx = \xx'$. The Robinson congruence \cite{pe1} is defined if we fix $\xx$ and  let ${\xx}^{\prime}$ vary.

Let $f$ be a primitive idempotent (PI) of $\CC\ot\cle\simeq\cl_{4,1}$ and $f_\pm:=\frac{1}{2}(1+\ee_3)$ be PIs of $\clt$.
Since the Dirac spinor $\psi$ is an element of the 
 ideal $(\CC\ot\cle)f \simeq \cle^+\simeq\clt \simeq \clt f_+ \oplus \clt f_-$, 
$\psi$ indeed consists, as well-known, of the direct sum of two Weyl spinors\footnote{The four types
 (dotted covariant, undotted covariant, dotted contravariant
and undotted contravariant) of algebraic Weyl spinors are indeed elements of the respective minimal lateral ideals $\clt f_-$, $f_+\clt$, $f_-\clt$ 
and $\clt f_+$ of the Pauli algebra $\cl_{3,0}$ \cite{rocha,wal1,wal2}.}.

Given a paravector $ x = x^0 + x^AE_A \in \RR\op\RR^{4,1}\hookrightarrow\cl_{4,1}$ 
define $\chi = xE_4\in\bigoplus_{k = 0}^2 \la^k(\RR^{4,1})$.

Now we define the twistor as an algebraic spinor  
$\chi \frac{1}{2}(1-i\g_5)Uf \in (\CC\otimes\cle)f \simeq \clt$, where
$U$ is a Clifford multivector and so $Uf$ is a Dirac spinor. 
The term\footnote{In order to get a clear correspondence between our formalism and the Keller index-free formulation of twistors,
by abuse of notation we adopt the same symbols to describe the DCWS.} 
$ \Pi:=\frac{1}{2}(1 - i\g_5)Uf = {0\choose \xi} \in\me(1 - i\g_5)(\CC\ot\cl_{1,3})$ is a DCWS. If we take again a basis
 $\{E_A\}$ of $\cl_{4,1}$ and a basis $\{\g_\mu\}$ of $\cle$, the isomorphism $\cl_{4,1}\simeq\CC\ot\cle$
 explicitly given by $E_0 = i\g_{0},\; E_1 = \g_{10},\; E_2 = \g_{20}, \; E_3 = \g_{30}$ and  $E_4 = \g_5\g_0 = -\g_{123}$
 is useful to prove 
the correspondence of this alternative formulation with eq.(\ref{twistor}), and so, with a geometric algebra index-free
version of the  Penrose classical twistor formalism, by eq.(\ref{twistor}). Indeed,
\beq
\chi \Pi &=&  (x^0E_4 + \ap^0E_0E_4 + x^1E_1E_4 + x^2E_2E_4 + x^3E_3E_4 + \ap^4)\Pi\nonumber\\
      &=&  x^0(-i\g_0\Pi) + x^k(\g_k\g_0)(-i\g_0\Pi) + \ap^0(i\g_0)(-i\g_0\Pi) + \ap^4\Pi\nonumber\\
      &=& (1 + \g_5\xx)\Pi =  {i {\vec{x}}\xi\choose\xi}.
     \eeq
\noi  

 The incidence relation determines a spacetime manifold point if we take $
J_{{\bar{\chi}}\chi} := {\ol{xE_4U}}xE_4U = -{\ol U}E_4 {\ol x}x E_4 U = 0,$
since the paravector $x\in \RR\oplus\RR^{4,1}$ is in the Klein absolute ($x{\ol{x}} = 0$). 
\subsection{Flagpoles and twistors from pure pinors and spinors}
A generalized flagpole is given by the 2-form
  $G = \me (i\mmu\tilde\mmu - i\mmu_C\tilde\mmu_C)$ \cite{bt}, where $\mmu_C$ is the charge conjugation of the pure spinor $\mmu$.
Given a real vector $p =  <i\mmu\mmu_C>_1$, corresponding  (modulo a real scalar) to
a family of coplanar vectors determining the generalized flagpole, let $\omega$ be an element of a maximal totally isotropic subspace of $V$
 such that 
$\omega \mmu^C = \mmu$, $\omega\mmu = 0$ and $\{\omega,\omega^*\}=0$. It can be shown that $G = \exp(i\theta) p \om + \exp(-i\theta) p \om^*$
and $F:=G\big|_{\theta = 0} = p(\om + \om^*) = {\rm Re}\;(i\mmu\tilde\mmu)$ is the Penrose flagpole \cite{pe2,bt}.

Now, from the well-known  correspondence between pure {\bf pinors} and the group O($2n$)/U($n$) \cite{harv}, it is possible to  adapt   
the proof of this correspondence, in order to establish the natural correspondence between pure {\bf spinors}, twistors and the group SO($2n$)/U($n$).
 
By definition, a spinor $\mmu$  is said to be \emph{pure} \cite{cartan,chev} if  the set 
$\Xi_\mmu := \{\ap\in \CC^{2n} \, :\, \ap(\mmu) = 0\}$
 has complex dimension $n$. Besides, the natural map from a pure spinor 
$\mmu$ to  $\Xi_\mmu$ induces an equivariant isomorphism 
from the algebra of  pure spinors ({\rm mod} $\CC^*$) to the set $\Xi_\CC$ of all $n$ dimensional totally null subspaces 
of $\CC^{2n}$. Now the well-known result proved in \cite{harv}, asserting that 
$\Xi_\CC \simeq {\rm O}(2n)/{\rm U}(n)$,  permits to 
link the pure spinors formulation to twistors. Indeed, the product of pure spinors 
is directly related to $n$-dimensional complex planes \cite{b4}, 
which are invariant (mod U(1)) under U($n$) actions. Thus 
it is possible, at least in even dimensions, to identify (via projective pure spinors) 
a twistor with an element of the group SO($2n$)/U($n$). Berkovits emphasizes this identification \cite{b4}.
In particular, twistors in four and six  dimensions are respectively elements of 
${{\rm SO(4)}}/{\rm U(2)}\simeq\CC\PP^1$ and ${\rm SO(6)}/{\rm U(3)} \simeq \CC\PP^3$.  
The investigation about an analogous mathematical structure and the physical implications of 
identifying twistors with elements of SO($2n$)/U($n$) is presented in \cite{b4}.  
\section{Twistors and graded exceptional structures}
It is well-known that it is possible, at  least in three, four, six  and ten dimensions, to construct a null vector
from spinors. In string twistor formulations some manifolds can be identified with the set of all spinors corresponding to the
same null vector, where in a particular case the homogeneous space SO(9)/G$_2$ arises \cite{ced1}.
Twistors are also an useful tool for the investigation of harmonic maps, as from the Calabi-Penrose twistor fibration $\CC\PP^3 \rightarrow S^4$ \cite{teteia}.
The deep relation between twistors and exceptional structures is illustrated in the classification of compact homogeneous
quaternionic-K\"ahler manifolds, the so-called \emph{Wolf spaces} \cite{alex,wolf}. 
The Wolf spaces associated with exceptional Lie algebras are  
{\rm E$_6$}/{\rm SU(6)$\times$Sp(1)}, {\rm E$_7$}/{\rm Spin(12)$\times$Sp(1)}, {\rm E$_8$}/{\rm E$_7 \times$Sp(1)}, {\rm F$_4$}/{\rm Sp(3)$\times$Sp(1)} 
and G$_2$/SO(4).
More comments concerning such structures are beyond the scope of the present paper (see \cite{alex,wolf,guna}).
It is also worth pointing out that the widespreading applications of exceptional structures in modern theoretical physics
give new possibilities for future advances. Since Kaplansky and Ka$\check{{\rm c}}$ extended, to $\ZZ_2$-graded algebras, 
the classification of finite-dimensional simple Lie algebras due to Cartan, superalgebras and supergeometry 
are being applied to  mathematical-physics, and the development of other gradings in exceptional Lie algebras \cite{cris} 
is a promisor field of research, with possible applications in mathematical-physics.

\section{Concluding remarks}
We presented   twistors in Minkowski spacetime as algebraic spinors associated to 
 $\mathbb{C}\otimes C\ell_{1,3}$,  using the paravector model, which was also used to describe all the conformal maps 
as the action of  twisted adjoint representations on paravectors of the Clifford algebra over $\RR^{4,1}$.
The identification of the twistor  identified  with SO($2n$)/U($n$) is obtained, from the complex structure based on pure 
 spinors formalism. As particular cases, twistors in four dimensions are elements of SO(4) modulo the
 double covering of  electroweak group SU(2)$\times$U(1), and in six dimensions twistors are elements of SO(6) modulo the double covering of the group SU(3)$\times$U(1).

\end{document}